# Fibrous thermoresponsive Janus membranes for directional vapor transport

Anupama Sargur Ranganath[a], Avinash Baji[a,b], Giuseppino Fortunato[c], René M. Rossi[c]

[a]*Pillar of Engineering Product Development, Singapore University of Technology and Design, Singapore 487372*

[b]*Manufacturing Engineering, LA TROBE University, Melbourne Victoria 3086, Australia*

[c]*Empa, Swiss Federal Laboratories for Materials Science and Technology, Laboratory for Biomimetic Membranes and Textiles, CH-9014 St. Gallen, Switzerland*

## Abstract

Wearing comfort of apparel is highly dependent on moisture management and the respective transport properties of the textiles. In today's used textiles, water vapor transmission (WVT) depends primarily on the porosity and the wettability of the clothing layer next to the skin and is not adapting or responsive to environmental conditions. The WVT is inevitably the same from both sides of the membrane. We propose a novel approach in this study by developing a thermoresponsive Janus membrane using electrospinning procedures. We targeted a membrane as a bilayer composite structure using polyvinylidene fluoride (PVDF) as one layer and a blend of PVDF and thermoresponsive poly-n-isopropyl acrylamide (PNIPAM) as the second layer changing wettability properties in the range of physiological temperatures. Tailored electrospinning conditions led to a self-standing membrane incorporating fiber diameters of 400nm and porosities of 50% for both layers within the Janus membrane. The WVT studies revealed that the combined effects of the Janus membrane's directional wettability and the temperature-responsive property results in temperature-dependent vapor transport. The results show that the membrane offers minimum resistance to WVT when the PVDF side faces the skin, which depicts the side with high humidity over a range of temperatures. However, the same membrane shows a temperature-dependent WVT behavior when the blend side faces the skin. From a room temperature of 25 °C to an elevated temperature of 35 °C, there is a significant increase in the membrane's resistance to WVT. This behavior is attributed to the combined effect of the Janus construct and the thermoresponsive property.
This temperature-controlled differential vapor transport offers ways to adapt vapor transport independence of environmental conditions leading to enhanced wearing comfort and performance to be applied in fields such as apparel or the packaging industry.

## Introduction

Tailored protective clothing is one of the oldest but actively researched fields for improving textiles' performance and comfort of textiles.[1, 2] The practical end-use include firefighter protection clothing, diver's, and space suit as well as raincoats, etc., drive the research in protective clothing. Typically, a wearer expects the garment to be functional and comfortable for the required end-use.

Sweat and heat transport of apparel describe the thermal comfort aspect,[3] which is a tradeoff between the performance and comfort properties of the clothing[4]. For example, rain ponchos have one of the best waterproof abilities as they are impermeable to water but uncomfortable to wear as the sweat cannot diffuse. Modern apparels employ a combination of novel chemistry and material structure to attain performance and comfort.[5-9] Even though the performance levels have improved from what it was decades ago, the comfort aspect still depends on the surrounding environment. Other than the touch, which is a qualitative factor, sweat transport, measured by water vapor transmission (WVT) across the fabric, is considered a quantitative measure of comfort in apparel. In a practical scenario, sweat transmission combines water and vapor transport, depending on the person's activity level. Liquid sweat transmission is required during a person's high activity level. In contrast, vapor transmission is needed during all activity levels of a person and is considered a measure of the fabric's comfort. It is often reported as a measure of comfort for the new material systems.[10-12]

WVT is primarily driven by the partial vapor pressure difference across the membrane and usually follows Fick's law of diffusion when a system is in a steady state. The temperature and humidity of the local environment govern the partial vapor pressure as expressed in Eq. 1. Therefore, WVT is achieved much better in arid [13] compared to humid regions.

$$p_w = \frac{H*6.11*e^{(17.67*\frac{T}{T+243.6})}}{100} \quad (1)$$

Free water vapor diffusion is often insufficient to increase comfort in humid regions, and forced ventilation is required to pump out the sweat. Attaching devices to circulate the air to remove sweat is cumbersome and practically inconvenient. Therefore, research on a material that can steer, adapt, and respond to environmental changes and pump out liquid sweat without external support is essential.

Electrospun fibrous membranes exhibit good water vapor permeability and wind resistance. Gibson *et al.* reported that the water vapor quickly diffuses out through the electrospun membrane due to the large porosity of the electrospun membrane. On the other side, the large surface area of the nanofibrous layer resists convective wind flow.[14] Furthermore, to improve the performance, electrospun membranes from selected polymers such as polyacrylonitrile (PAN) and polyurethane (PU) were modified to enhance their tensile stress and breaking elongation %, waterproof ability with tailored vapor permeability.[15-17] Selected studies evaluated the vapor permeability of multilayered membranes[18], a combination of electrospun and woven textiles[19], and electrospinning of hydrophilic/hydrophobic layers.[20] Multilayered membrane samples investigated by Mukhopadhyay *et al.*[18] showed that the water vapor transmittance is dependent on the porosity and pore size of the middle layer of polyester fleece/polyester spacer, even when the porosity of the innermost and outermost layers is constant in all the tested samples. Therefore, the multilayer system, which had a highly porous middle layer, exhibited larger WVT due to increased overall porosity. Investigation of the woven fabric coated with an electrospun fibrous mat by Bagherzadeh *et al.* [19] showed that the electrospun layer did not impede the water vapor permeability of the woven fabric. In another investigation of the multilayer electrospun membrane by Gorji *et al.*[20] they revealed the effect of incorporating graphene oxide in the hydrophilic matrix, which is layered adjacent to a polyurethane fibrous membrane. The water vapor permeability reduced with higher hydrophilic layer's weight. Increasing graphene oxide content from 0.1 to 0.4% in the acrylamide-based hydrophilic polymer, reduced the water solubility of the polymer and consequently increasing the water vapor permeability. However, studies on multi-layer systems did not investigate the direction of the vapor transmission across the thickness of the membranes. As previous studies revealed that the vapor diffusion through the microporous membrane is porosity dependent, the vapor transmission from either side of the membrane along its thickness is assumed to be constant.

Introducing heterogeneity in the membrane chemistry across the thickness, without altering the porosity, exhibits directionality in the membrane's properties. E.g., combining hydrophobic and hydrophilic material in layers within a single membrane exhibits directional water flow. Such membranes, with faces of different chemistry, are termed Janus membranes[21, 22]. These membranes have attracted high attention from researchers. In one such system, the water drops flow from the hydrophobic side to the hydrophilic side but not from the hydrophilic to the hydrophobic side. The flow from the hydrophobic side is due to the hydrophilic layer underneath the hydrophobic, which pulls the droplet across the membrane. The Laplace pressure difference, along with the thickness of the membrane, explains this mechanism.[23]

Another class of materials, i.e., environmentally responsive polymers, such as poly N-(isopropylacrylamide), are termed 'smart' due to their switchable properties in response to environmental cues.[24] Figure 1 shows the change in molecular conformation of PNIPAM concerning the environmental temperature. At room temperature, the carbonyl and amide groups of PNIPAM are exposed to form a hydrogen bond with the surrounding water vapor. However, at elevated temperatures, the hydrogen bonds break and cause intramolecular bonding between carbonyl and amide groups from adjacent monomer units. This coil conformation is relatively hydrophobic compared to the extended conformation at room temperature, being hydrophilic.[25-27]

The PNIPAM-based hydrogel can be coated on textiles like cotton or nylon 6 fabrics and exhibit thermoresponsive behavior. The WVT studies by Stular et al., and Verbič et al. show less vapor transmission at ambient temperature in comparison with WVT at an elevated temperature of 40 °C. The swelling of PNIPAM reduces the porosity and, in turn, reduces the vapor transmission at ambient temperature

Independent research on responsive materials,[24, 28, 29] and Janus constructs[30-32] shows high potential for smart and self-sustaining systems with directional liquid transport. Our current study shows the result of combining responsive material such as PNIPAM in a Janus construct. PNIPAM is blended with PVDF in a 25:75 wt% ratio to minimize the effect of swelling and retain the thermoresponsive behavior. The blend and pristine PVDF are electrospun in layers to obtain a Janus construct. Consequently, the electrospun Janus membrane has PVDF on one face and a blend on the other. We use two independent experimental approaches to assess and confirm the WVT performance of the membrane. For the first time in literature, we show that WVT is preferentially more in one direction within a given membrane. The directionality is plausibly due to the 'passive pumping' action of the Janus membrane, combined with the thermoresponsive property of the hydrophilic layer.

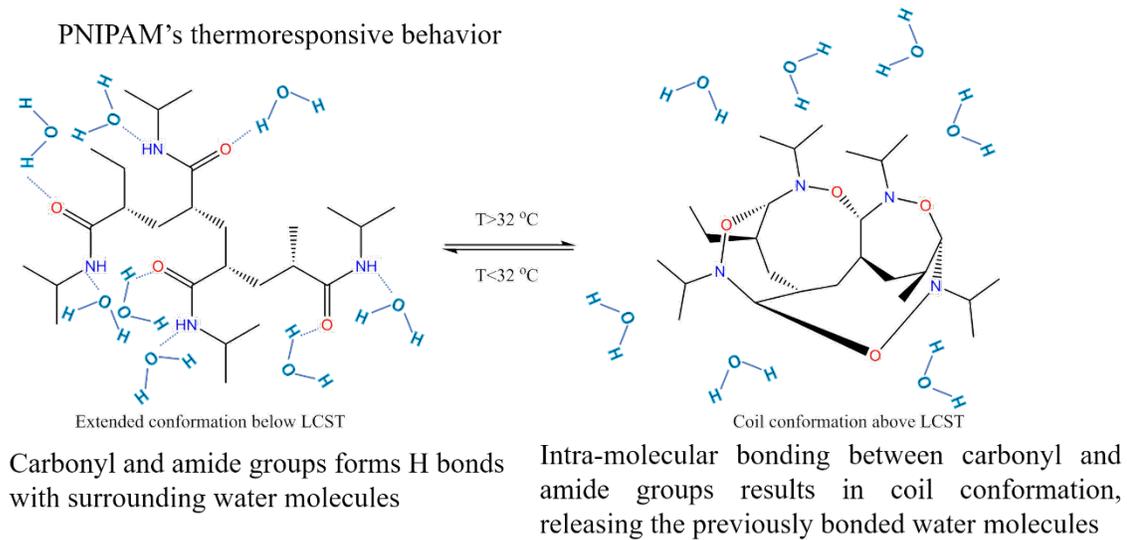

*Figure 1: Shows the change in the molecular conformation of poly N-(isopropyl acrylamide) in response to the change in the environmental temperature.*

## Materials and methods

### Materials
PVDF pellets with a molecular weight of 180000 g/mol and 530000 g/mol powder were procured from Sigma-Aldrich, Switzerland. PNIPAM powder with a molecular weight of 300000 g/mol was purchased from Scientific Polymer Products Inc. N, N-dimethylformamide (DMF, 99.5 %) from Sigma-Aldrich Switzerland.

### Methods

#### Membrane fabrication
The PVDF solutions were prepared by dissolving 33 wt% of PVDF (180000 g/mol) pellets in DMF at 60 °C overnight. The blend solutions were prepared by dissolving 18 wt% of the polymer mixture, i.e., PVDF (530000 g/mol) and PNIPAM (300000 g/mol) in 75/25 w/w ratio in DMF, and magnetically stirred at 60 °C on a hot plate overnight and subsequently cooled to room temperature before electrospinning.

#### Needle-based electrospinning

PVDF solution or the blend solution was loaded in a plastic syringe with a 21G blunt needle (0.8 mm inner diameter). The flow rate of the polymer solution was set to $0.5 ml*h^{-1}$ using a flow pump. The needle tip was connected to a voltage of 10kV and the collector plate to a voltage of -5kV. The working distance between the needle and the flat plate aluminum collector was 12 cm. The aluminum collector was covered with silicone paper for easy peeling off the electrospun membranes.

Needle-less electrospinning (Nanospider[TM])

Nanospider[TM] (Elmarco, Czech Republic) is a needle-less electrospinning technology with upscaling capability. Figure 1 shows a schematic of the procedure. The following spinning parameters were used for homogenous spinning: The vertical gap between the two wires was 25 cm, wherein the top wire was applied with a voltage of +60 kV and the bottom wire with -10 kV. The traversing carriage, with a speed of 270 mm/s on the bottom wire, housed a pinhole of 0.5 mm, which controls the volume of polymer solution trailed on the bottom wire. The collector paper moving at a speed of 18 mm/min, is placed right below the top wire

After 20 minutes of electrospinning, the paper position is unrolled to the starting point to electrospin in the same area. Five such repetitions build a thick and wide membrane with a surface area of 500 mm$^2$. The fabrication of the blend and PVDF layer followed the same procedure. Four Janus membranes were prepared and tested for water-vapor resistances using the sweating guarded hotplate, elaborated in the following sections.

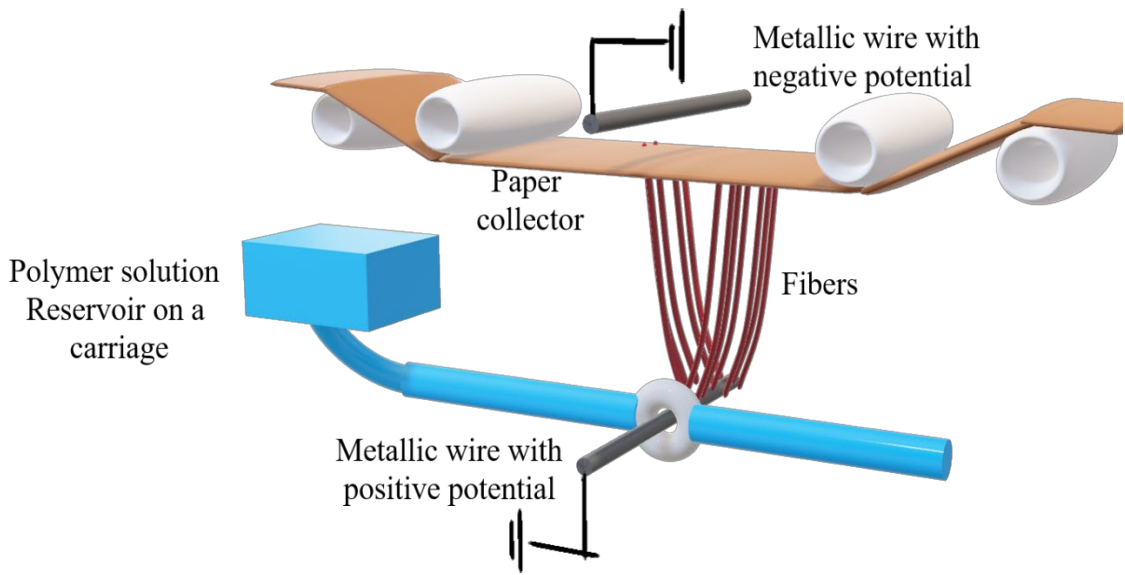

*Figure 2: Needle-less electrospinning setup used to develop sub-micron-sized fibers at the pilot-scale level.*
*A carriage with polymer solution traverses the bottom metallic wire (+60 kV) and leaves a trail of the solution droplets on this wire. The potential difference between the wires draws the fibers from these droplets. The silicone paper collects the fibers placed right below the top metallic wire (-10 kV).*

Material characterization

The viscosity of the polymer solutions was evaluated using a Physica MCR301 Rheometer (Anton P, Graz, Austria) with a plate-cone geometry. The shear viscosity of the polymer solutions was assessed as a function of the shear rate. The spinning solution's electrical conductivity was measured using Metrohm 660, Switzerland. The electrospun membranes were visually examined using scanning electron microscopy (SEM) by a Hitachi S-4800, Hitachi High-Technologies, USA & Canada) using 2 kV accelerating voltage and 10 mA current flow. Before the SEM measurements, the samples were sputtered with 8nm of Au-Pd to increase the conductivity using a sputter coater, Leica EM ACE600, Leica Microsystems, Germany.

Sweating guarded hotplate

The sweating guarded-hotplate to determine the resistance of the membrane to WVT follows ISO 11092[33]. It is often referred to as the "skin model" as it simulates the heat and vapor transfer processes next to the skin. Primarily, it consists of an electrically heated porous plate to simulate the thermoregulation model of the skin (Figure 2). Heat loss is avoided by using the guard underneath and on both sides of the hot plate. At the same time, the guards are heated to the same temperature as the porous plate. The water-circulating system feeds the heated plate to produce the vapor by evaporation. The system underneath the plate measures the heating power required to maintain the temperature of the plate. The measurement is carried out in a controlled environment as it involves temperature, relative humidity, and wind speed combinations.

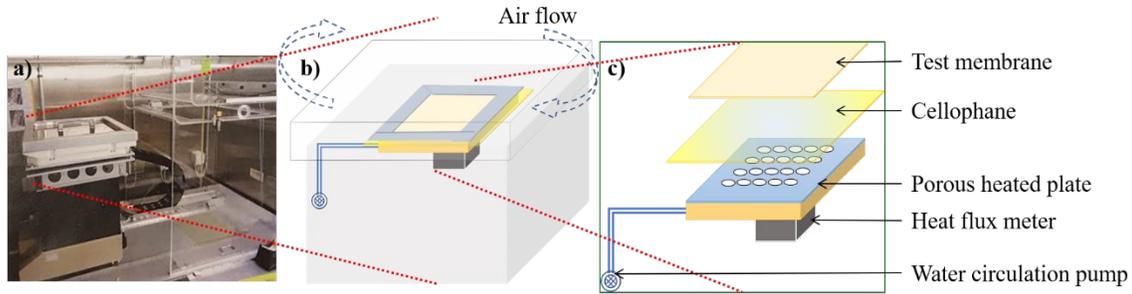

*Figure 3: Schematic of the sweating hot plate instrument. a) is the photographic image of the device. b) shows the airflow tangential to the mounted test sample. c) shows the parts in layers. The circulation pump supplies the water to the electrically heated and porous plate for evaporation. Cellophane is a waterproof but vapor permeable layer to transmit the evaporating vapor from the plate. The test membrane lays flat on the cellophane covered heated plate and is held in place by a frame on all sides.*

The membrane is placed on an electrically heated plate, covered with a saturated cellophane sheet permeable to vapor but impermeable to water. Air is tangentially blown across the membrane surface to maintain a constant vapor pressure gradient. This setup permits only the vapor from the plate to pass through the fibrous membrane and prevents the water from wetting the fibrous membrane. The heat flux required to maintain the saturated vapor pressure is a measure of the membrane's resistance to vapor permeability. The expression for the resistance to vapor permeability is as follows:

$$R_{et} = \frac{[\rho_m - \rho_a]}{H - \Delta H_e} \qquad (2)$$

$R_{et}$: Water-vapor resistance in $m^2 Pa/W$

$\rho_m$: The saturation water-vapor partial pressure in Pascal (Pa), at the surface of the heated plate at a temperature in °C

$\rho_a$: The water-vapor partial pressure in Pa of the air in the test enclosure with air temperature in °C

$H$: Heating power supplied to the measuring unit in W

$\Delta H_e$: Baseline error correction term for heating power for the measurement of water-vapor resistance $R_{et}$, used as a reference value for ambient conditions.

The boundary layer resistance of the cellophane layers is the baseline measurement of the system. The software deducts the resistance of the boundary layer from the experiment results for subsequent measurements. Thus, the instrument directly calculates the resistance offered by the fibrous membranes.

OptiCal double-chamber method

OptiCal from Michell Instruments is a relative humidity (RH) and temperature calibrator that uses an optical sensor for high-precision measurements. It is used to design a setup that determines the WVT of the membrane. Figure 4 shows the setup schematic, which consists of a temperature-controlled chamber with a sealed container for the test membrane. A reservoir draws water from a tube placed on the weighing scale to maintain its level to ensure a controlled RH. A climatic chamber with controlled temperature and RH houses the entire setup. A computer-connected software controls the environmental conditions. In parallel, another software records the weighing scale measurements, i.e., the weight reduction due to water flow into the OptiCal chamber is directly associated with vapor diffusion through the membrane

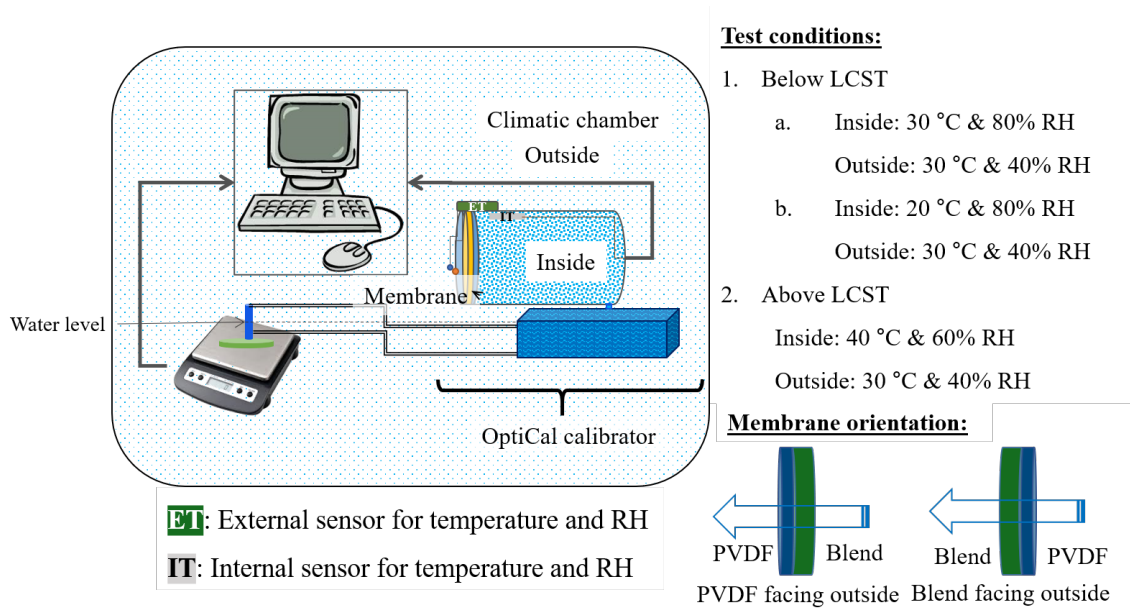

*Figure 4: Schematic of the double chamber setup.*

Measurement of water vapor permeability started after the stabilization of the environmental conditions. The water vapor passes through the membrane (∅ = 0.06 m) from the OptiCal chamber. As the RH in the OptiCal chamber drops, water from the reservoir is evaporated to maintain the desired RH. The water from the tube on the scale flows to the reservoir to maintain the water level in the reservoir. The reduced amount of water from the tube is weighed by scale, and the weight loss is recorded in real-time. Water vapor permeance is the weight loss over a defined period for a unit partial vapor pressure difference across the membrane. The environmental conditions between the inside and outside instruments govern this partial vapor pressure difference. An external RH and temperature sensor from MSR® with a built-in data logger measured the test conditions outside the membrane surface. 0.05" thermocouple wires were embedded into the fibrous membranes to record the actual temperature at the surface and the interface of two layers of the Janus membrane. The expression for the partial vapor pressure as a function of temperature and humidity is given by following equations[34, 35].

$$p_w = \frac{H * 6.11 * e^{(17.67 * \frac{T}{T+243.6})}}{100} \tag{3}$$

$p_w$ is the partial pressure, H is the humidity in %, and T is the temperature in °C.

$$WVP = \frac{W}{t * (\rho_{in} - \rho_{out}) * A} \tag{4}$$

WVP is the water vapor permeability in $g/(h * m^2 * mbar)$, W is the water loss in grams, t is the time in hours, $p_{w\,in} - p_{w\,out}$ is the water vapor partial pressure difference in mbar between inside and outside conditions, A is the membrane area in $m^2$.

Figure 4 lists the testing condition of the experiments. It was possible to maintain isothermal conditions in the system below the lower critical solution temperature (LCST). As the recommended operating temperature of the OptiCal was less than 30 °C, it was not possible to maintain isothermal conditions above LCST. However, to ensure that the membrane is above LCST and to minimize the thermal gradients, the constant temperature condition of 30 °C and 40 °C are maintained outside and inside the chamber, respectively.

## Results and discussion

The electrospun membrane with Janus construct was fabricated using a needle-less and needle-based electrospinning setup (see Table 1). The needle-based electrospinning setup allowed us to incorporate the thermocouples between the layers and just below the surface. Therefore, the precise measurement of surface temperature and that between the layers was precisely measured for the double-chamber method. These measured temperatures were used to calculate the vapor pressure gradient across the membrane.

The PVDF solution was electrospun on top of the electrospun blend membrane to produce a two-layered Janus construct. The blend solution with a concentration of 16 wt% had a shear viscosity of 2.38 Pa.s under a shear of 1/10 s and a conductivity of 5.16 μs/cm. Similarly, the shear viscosity of the PVDF solution is 1.2 Pa.s and a conductivity of 32 μs/cm. The SEM examination shows a smooth fiber morphology with a diameter of 0.2-

0.4 μm for needle-less electrospun fibers and 0.2-0.6 μm for needle based electrospun fibers. These values indicate no dimensional difference between the fibrous web produced by needle based or needle-less electrospinning methods. The specific weight of the Janus membranes is 30-40 GSM, a lightweight fabric category. However, in electrospun or thin-film membranes, this weight range indicates a heavyweight membrane suitable for practical applications[36].

Table 1: Polymeric solution parameters and their corresponding electrospun membrane properties.

| Polymer (MW, Da) | Wt% (w/w) | Shear viscosity Pa.s, (at 1/10 s) | Conductivity, μs/cm | Duration, no of cycles of 20min each | Fiber diameter, nm | Thickness, μm | Porosity, % |
|---|---|---|---|---|---|---|---|
| PVDF(530K)/PNIPAM (330K)(75/25) | 18 | 2.4 | 5.16 | 5 | 502 ± 193.8 | 103.3 ± 5.4 | 49.7 ± 1.8 |
| PVDF (180K) | 30 | 1.2 | 32 | 3 | 139.2 ± 76.2 | 67.7 ± 4.7 | 49.6 ± 3.3 |

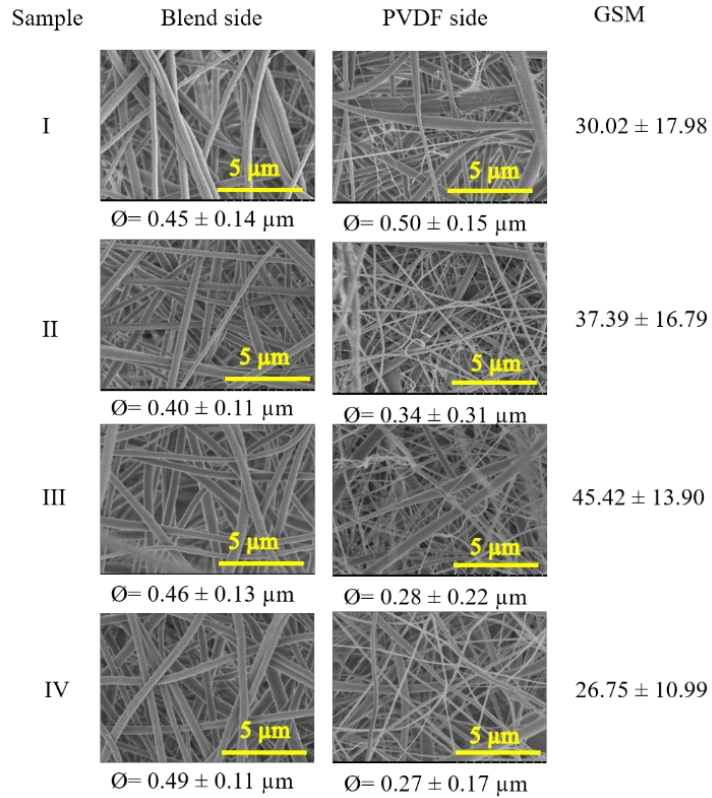

| Sample | Blend side | PVDF side | GSM |
|---|---|---|---|
| I | Ø= 0.45 ± 0.14 μm | Ø= 0.50 ± 0.15 μm | 30.02 ± 17.98 |
| II | Ø= 0.40 ± 0.11 μm | Ø= 0.34 ± 0.31 μm | 37.39 ± 16.79 |
| III | Ø= 0.46 ± 0.13 μm | Ø= 0.28 ± 0.22 μm | 45.42 ± 13.90 |
| IV | Ø= 0.49 ± 0.11 μm | Ø= 0.27 ± 0.17 μm | 26.75 ± 10.99 |

*Figure 5: SEM micrographs of the four Janus membranes prepared using needle-less electrospinning. Blend fibers are relatively more uniform in comparison to PVDF fibers, which have a bimodal distribution of fiber diameter. GSM refers to the membrane weight in grams per square meter. I to IV are the Janus samples electrospun for measuring WVT*

Figure 6 shows the surface elemental composition by X-ray photoelectron spectroscopy (XPS) for PNIPAM, PVDF, and their blend. Comparing the blend with respective pristine counterparts reveals that the blend surface is enriched with Nitrogen and Oxygen, which is like PNIPAM fibers. The XPS results confirm the observations from our previous study on the thermoresponsive wettability of PNIPAM/PVDF blends fabricated using needle-based electrospinning[29]. The Thermal characterization using DSC and TGA suggested the phase separation of PNIPAM and PVDF during electrospinning. At the same time, the wettability switch observed by contact angle measurement at room temperature and elevated temperature suggested PNIPAM enriching the fiber surface[29].

A comparison of density and solubility parameters in Table 2 favors the dissolution of PNIPAM in DMF over PVDF. Therefore, the evaporation of DMF during the electrospinning process supports the migration of PNIPAM to the fiber surface, lasting longer in solution. Our previous study on this blend suggests miscibility when the PNIPAM content is 50 wt% or above[29]. However, when the PNIPAM content is 25 wt% or below, PVDF and PNIPAM phases separate, enhancing the migration of PNIPAM to the fiber surface during the electrospinning process.

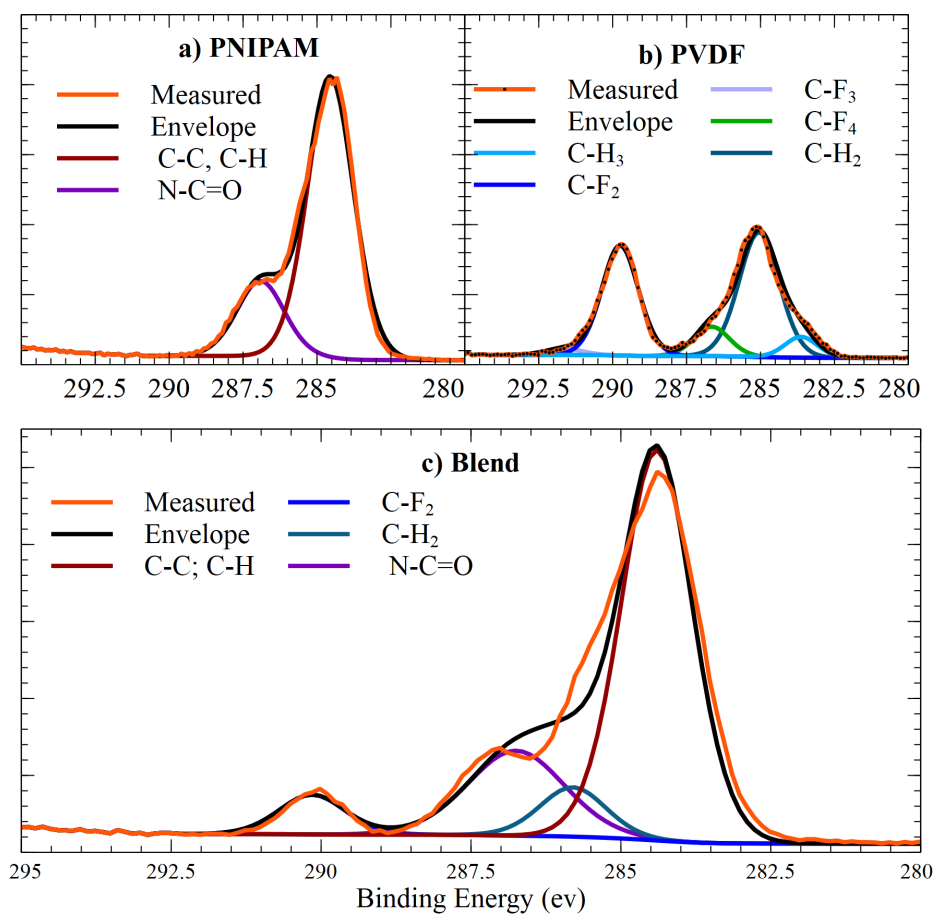

| Elements/transitions | XPS surface composition, at.% ||||
|---|---|---|---|---|
|  | PVDF | PNIPAM | BLEND | BLEND after three months in water |
| Carbon C1s | 51.1 | 75.9 | 71.8 | 71.8 |
| Nitrogen N1s | - | 10.9 | 8.7 | 9.4 |
| Oxygen O1s | 3.0 | 13.2 | 11.8 | 9.6 |
| Fluorine F1s | 45.9 | - | 7.8 | 9.3 |

Figure 6: XPS graphs of PNIPAM, PVDF, and the blend of PVDF/PNIPAM (75/25, w/w)

| Table 2: Polymer properties |||||
|---|---|---|---|---|
| **Polymer** | **Density** | **Solubility parameter δ2** | **Solubility parameter DMF δ1** | **Δδ (1-2)** |
| PNIPAM | 1.05 | 23.5 | 24.9 | 1.4 |
| PVDF | 1.68 | 17.5 | 24.9 | 7.4 |

Before water-vapor transmission experiments were performed, the fabricated membranes were conditioned for a day in an environment-controlled chamber (at test conditions).

The skin model (mimicked by the porous hot plate) measures the membrane's resistance to vapor permeability as a function of temperature, as shown in Figure 7. When the membrane is placed on the hot plate, the water vapor diffuses from the bottom to the top side of the Janus membrane (see Figure 7). The membrane's resistance to vapor diffusion is measured from both sides of the Janus membrane in a separate set of experiments to assess the influence of the wettability gradient within the membrane. This set of measurements is performed at five different temperatures to plot the membrane's resistance as a function of temperature. The water vapor resistance measurement removes the temperature bias on the WVT and is expected to be constant for the same fabric at different isothermal conditions.

When the membrane is placed on the hot plate with the blend side facing down, i.e., when vapor transmits from the blend to the PVDF side of the Janus membrane, there is an increased water vapor resistance at higher temperatures (see Figure 7). As the blend is thermoresponsive, it is hydrophilic (CA=10º) at a lower temperature range (<32 ºC) and hydrophobic (CA=120º) at a temperature higher than 32 ºC. As a result, the lower resistance is attributed to the blend's affinity to water vapor at a lower temperature. Similarly, the reduced affinity at elevated temperatures causes more significant resistance to vapor transmission. As the vapor transmits from the blend to the PVDF side, the hydrophobicity increases the membrane's thickness and consequently increases the membrane's resistance to vapor transmission.

The hydrophilic layer next to the PVDF layer supports the vapor transmission when the PVDF side faces the hot plate. At elevated temperatures, the resistance increases but is significantly lower than the resistance the membrane offers, with the blend facing the hot plate (Figure 7). Even though the blend is hydrophobic at elevated temperatures, it is less hydrophobic than PVDF. Therefore, when the vapor transmits from the PVDF to the blend side, the hydrophobicity reduces along with the thickness of the membrane, which favors the vapor transmission.

Based on the examination of the results, the Janus construct with PVDF facing the hot plate favors the vapor transmission at all investigated temperatures. This behavior is due to unchanging hygroscopic properties of the PVDF with temperature. Therefore, the thermoresponsive Janus membrane makes it possible to maintain active vapor transport irrespective of the outside temperature.

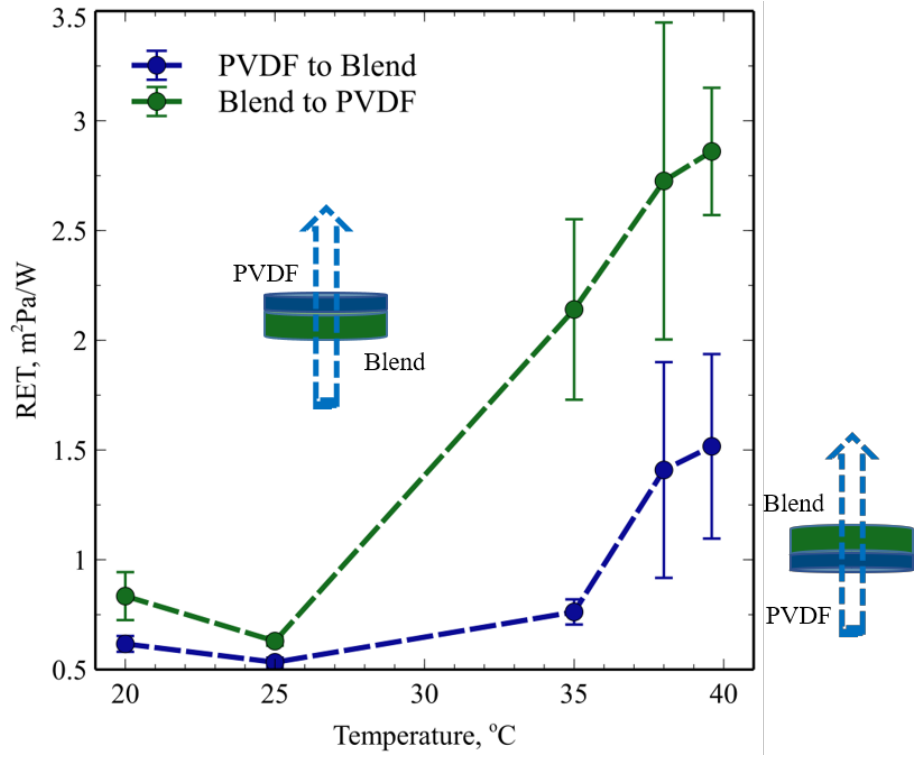

*Figure 7: Effect of Janus directionality on the resistance to water-vapor permeability through the membranes. The thermoresponsive property of the blend combined with the Janus structure offers higher resistance to WVT. The behavior is attributed to the moisture released by the blend layer at a higher temperature. As a result, when the blend faces the hot plate, it increases the boundary layer gap and consequently increases the resistance to water-vapor permeability.*

We measured the water vapor permeability using a double chamber method to verify the observed behavior. Needle-based electrospinning was used to fabricate the Janus membrane to incorporate the thermocouples between the layers and almost at the surface of the Blend layer (approximately 5sec of electrospinning). Figure 8 shows the Janus membrane with thermocouples on the sample holder to fit the mouth of the OptiCal chamber. The figure also shows the SEM micrographs of the blend and PVDF side of the Janus membrane incorporating a fiber diameter of 0.2-0.6 μm.

The sample holder plugs the mouth of the chamber such that one of the sides of the membrane faces outside the chamber and the other faces the inside chamber of the OptiCal chamber. The entire system and the membranes were conditioned at 20°C and 40% RH before carrying out below LCST. Before measurements above LCST, membranes were conditioned at 30°C and 40% RH, as mentioned in Figure 4.

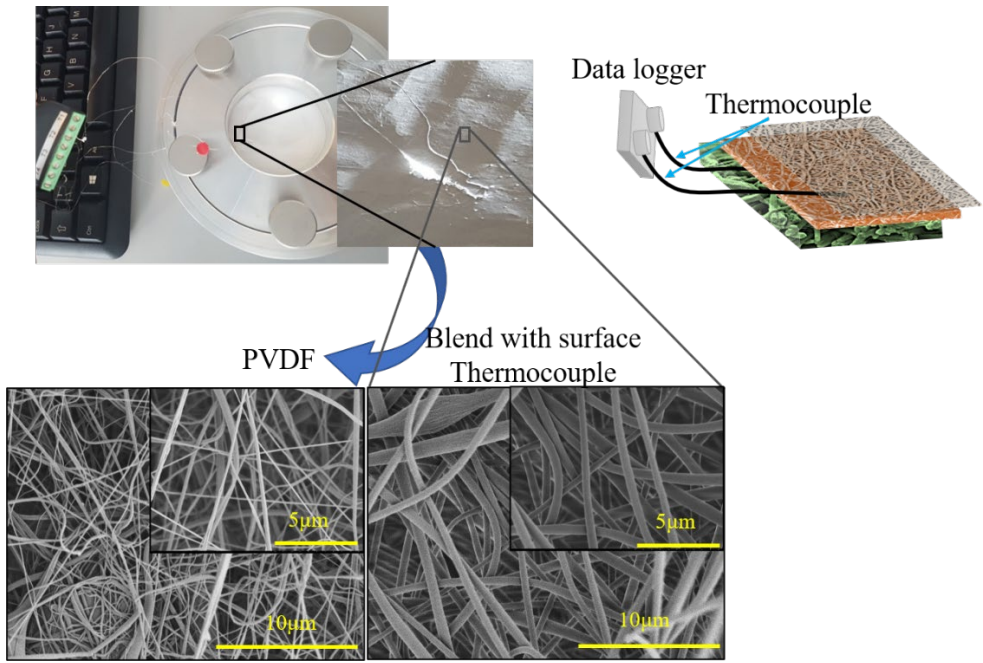

Figure 8: The top part shows the membrane with the thermocouples mounted (red arrow) on the sample holder that fits the mouth of the OptiCal instrument. The bottom section shows the SEM micrographs of the fibers from the PVDF and the blend side, respectively

Figure 9 shows the membranes' water vapor permeability as a temperature function. At a lower temperature of 20 °C, the permeabilities are comparable for both samples. However, the increasing vapor permeability with increasing temperature is predominantly due to increasing partial vapor pressure difference across the membrane[37]. Figure 10 plots the vapor permeability as a function of partial vapor pressure across the membrane.

PVDF being hydrophobic is expected to adsorb less moisture and transmit less than the unswelling hydrophilic membrane (Blend). However, interestingly, the vapor permeability from the PVDF to the blend side is significantly higher than from the blend to the PVDF side. Further, to isolate the membrane effects from the vapor pressure, Figure 10B shows the water vapor permeability per unit of partial pressure across the membrane. When the vapor transmits from the blend side, the permeability is constant, suggesting the vapor pressure difference is the primary driving force. However, vapor permeates significantly more (P-value =0.003, n=4) from the PVDF side, suggesting that the membrane's influence on vapor transmission increases with temperature. Therefore, the combined effect of the Janus constructs and the temperature-responsive property drives more vapor through the membrane from one direction over the other.

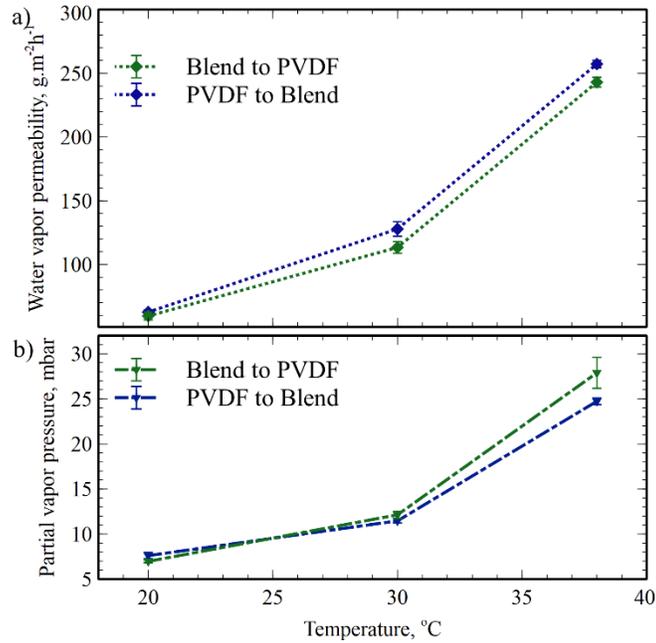

*Figure 9: a) shows the Janus membrane's water vapor permeability as a temperature function and compares the effect of the membrane's directionality. b) shows the partial vapor pressure difference across the membrane as a function of temperature. The water vapor permeability for PVDF to blend direction is significantly higher due to the additional partial pressure drop caused by the wettability gradient in the membrane.*

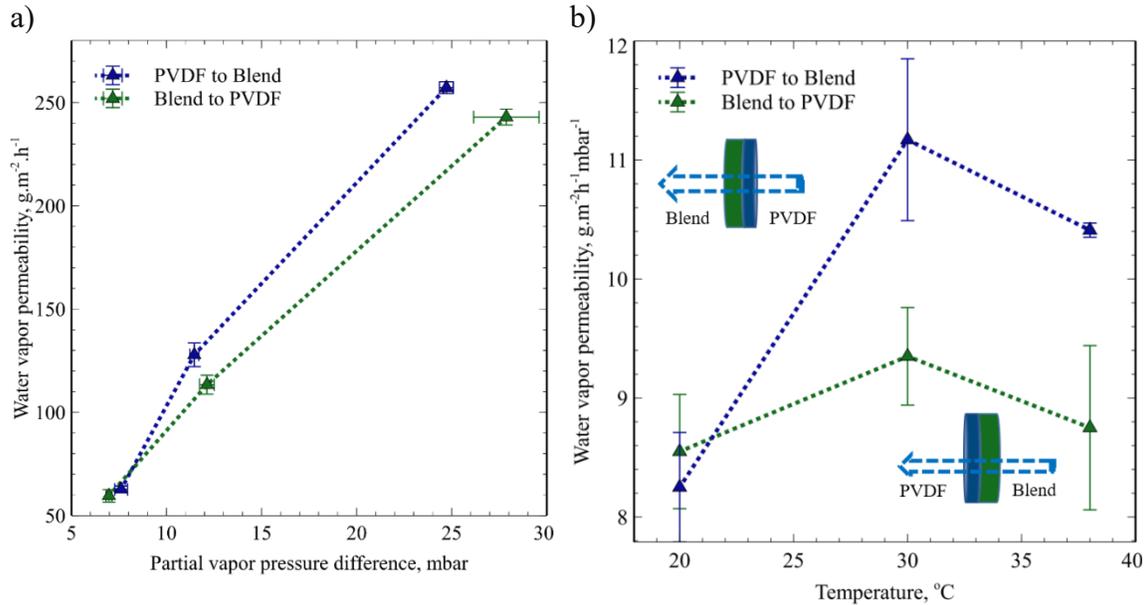

*Figure 10: a) Water vapor permeability as a function of the partial vapor pressure difference across the Janus membrane. Vapor permeability from the PVDF side is significantly higher than from the blend side due to the Janus structure favoring the vapor transmission from the PVDF to the blend side.*
*b) Water vapor permeability per unit of the partial vapor pressure difference across the Janus membrane as a function of temperature. The differences in the vapor permeability from the blend side to PVDF are statistically insignificant at all tested temperatures. The partial vapor pressure difference across the membrane is the primary driving force for transmitting the water vapor from the blend side. However, this permeability per unit of the pressure from the PVDF side increases significantly with temperature due to the Janus construct's combined effect and the blend's thermoresponsive property.*

Figure 10 shows the possible mechanism based on the obtained experimental results. Based on the prior hygroscopic measurements, pristine PNIPAM adsorbs 19 wt% of vapor at a temperature of 40 °C[38]. As the blend contains 25 wt% of PNIPAM and adsorbs water vapor in proportion to the PNIPAM content[29], which is at least 4 wt% of vapor at 40 °C. However, PVDF being hydrophobic, adsorbs less than 1% of vapor via Van der Waal's forces[29]. Therefore, at any given time during the experiment, the blend surface layer will hold more moisture (vapor molecules) ($C_B$), when compared with the PVDF surface layer ($C_P$).

When the PVDF side faces outside, the vapor transmits from the blend to the PVDF side of the membrane. The amount of vapor molecules available for evaporation is $C_P$ on the PVDF surface. Therefore, the vapor pressure gradient drives $C_P$ molecules through the membrane. As this concentration of vapor $C_P$ does not change with temperature, we have almost the same vapor permeability when transmitting from the blend to the PVDF side.

In the other scenario, with the blend side facing outside, the vapor transmits from the PVDF side to the blend side of the membrane. At equilibrium, the blend surface holds more vapor than the PVDF side. However, at 20 °C, most vapor molecules form hydrogen bonds with the amide (-N-H) and carbonyl (-C=O-) groups from PNIPAM. As a result, the concentration of vapor molecules ($C_B$) is a combination of bound vapor (T) and free vapor molecules (F). The vapor pressure gradient drives the free vapor molecules (F) through the membrane, which is comparable with $C_P$. Therefore, vapor permeability at temperatures below LCST is similar irrespective of the transmission direction.

Above LCST, due to the coil conformation of PNIPAM, all the bound vapor molecules are released and become free vapor molecules. When the vapor molecule $C_B$ evaporates, the vapor pressure gradient drives $C_B$ through the membrane. As $C_B \gg C_P$, the vapor permeability from PVDF to the blend side is higher than that from the flipped direction.

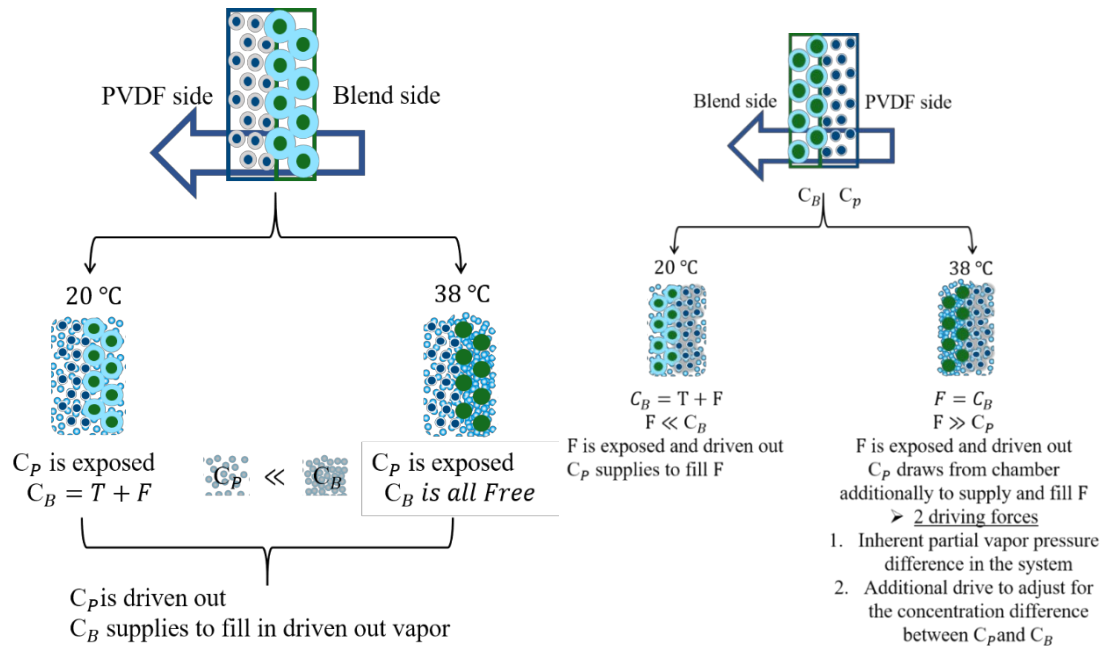

Figure 11: Illustrates the mechanism of water vapor permeability from the blend side to the PVDF side and vice-versa when driven by the partial vapor pressure difference.

# Conclusion

Electrospun thermoresponsive Janus membranes exhibit directional WVT. Herein, the vapor transmitted from the hydrophobic (PVDF) side to the hydrophilic (blend) is faster than in the opposite direction (hydrophilic to hydrophobic). The results from the indirect approach, i.e., via the sweating hot plate method, and from the direct approach, i.e., via a double-chamber method, complement each other. Based on the physical reasoning, we postulate that this mechanism is due to the combined effect of the temperature-responsive behavior of the Janus construct on vapor transmission. By complementing the experiment, numerical modeling can shed further insight into the physical processes, which results in directional vapor permeability.

These new results open pathways in membrane research and development, which is unique for liquid and gas transmission. The novelty not only contributes to the field of textiles, packaging, or filter systems.